\documentclass[journal,comsoc]{IEEEtran}
\usepackage[T1]{fontenc}

\usepackage{times}
\usepackage{soul}
\usepackage{url}

\usepackage{graphicx}
\usepackage{subfigure}
\usepackage{amsmath}
\usepackage{amsfonts}
\usepackage{algorithm}
\usepackage{algorithmic}
\usepackage{multirow}
\usepackage{booktabs}
\usepackage{lineno}
\usepackage{pifont}
\usepackage{color}
\usepackage{verbatim}
\usepackage{hyperref}

\graphicspath{ {images/} }
\ifCLASSINFOpdf
\else
\fi

\usepackage{amsmath}
\usepackage[cmintegrals]{newtxmath}
\begin{document}

\title{DeepPoison: Feature Transfer Based \\ Stealthy Poisoning Attack for DNNs}

\author{Jinyin~Chen,
        Longyuan~Zhang, Haibin~Zheng, Xueke~Wang
        and~Zhaoyan~Ming% <-this % stops a space
\IEEEcompsocitemizethanks{\IEEEcompsocthanksitem J. Chen is with the Institute of Cyberspace Security and the College of Information Engineering, Zhejiang University of Technology, Hangzhou, 310023, China
\IEEEcompsocthanksitem J. Chen, L. Zhang, H. Zheng and X. Wang are with the College of Information
Engineering, Zhejiang University of Technology, Hangzhou 310023, China.
\IEEEcompsocthanksitem Z. Ming is with the Institute of Computing Innovation, Zhejiang University, Hangzhou 310007, China.
\IEEEcompsocthanksitem Corresponding author: Zhaoyan Ming, e-mail: mingzhangyan@gmail.com
}% <-this % stops an unwanted space
\thanks{Manuscript received November 23, 2020;}}

% \markboth{IEEE TRANSACTIONS ON CIRCUITS AND SYSTEMS—II: EXPRESS BRIEFS}%
% {Chen \MakeLowercase{\textit{et al.}}: Bare Demo of IEEEtran.cls for Computer Society Journals}
%\markboth{ IEEE TRANSACTIONS ON CIRCUITS AND SYSTEMS—II: EXPRESS BRIEFS,~Vol.~XX, No.~X, XX~XXXX}%
%{Shell \MakeLowercase{\textit{et al.}}: Bare Demo of IEEEtran.cls for Computer Society Journals}

\maketitle

% \IEEEtitleabstractindextext{%
\begin{abstract}
Deep neural networks are susceptible to poisoning attacks by purposely polluted training data with specific triggers. \textcolor{black}{As existing episodes} mainly focused on attack success rate with \textcolor{black}{patch-based} samples, defense algorithms \textcolor{black}{can easily detect these poisoning samples}.  We propose DeepPoison, a novel adversarial network of one generator and two discriminators, to address this problem. Specifically, \textcolor{black}{the generator automatically extracts the target class'} hidden features and \textcolor{black}{embeds them} into benign training samples. One discriminator controls the ratio of the poisoning perturbation. \textcolor{black}{The other discriminator works as} the target model to testify the poisoning effects. The novelty of DeepPoison lies in that the generated poisoned training samples are \textcolor{black}{indistinguishable} from the benign ones by both defensive methods and manual visual inspection, and even benign test samples can achieve the attack.
Extensive experiments have shown that DeepPoison can achieve a state-of-the-art attack success rate, as high as 91.74\%, with only 7\% poisoned samples on \textcolor{black}{publicly available datasets} LFW and CASIA. Furthermore, we have experimented with  \textcolor{black}{high-performance} defense algorithms such as autodecoder defense and DBSCAN cluster detection and showed the resilience of DeepPoison.
\end{abstract}

% }

\maketitle
\IEEEpeerreviewmaketitle

\section{Introduction}\label{sec:introduction}

\IEEEPARstart{P}{oisoning} attack on a deep neural network (DNN) refers to an attack method that paralyzes the benign model or enables the compromised model to achieve the attacker's goal toward \textcolor{black}{specific} class labels. It usually injects backdoor to the model,  activated by a particular pattern in the testing samples. Poisoning attack has attracted attentions various applications, i.e., computer vision~\cite{krizhevsky2012imagenet,he2016deep}, speech signal processing~\cite{graves2013speech} and  natural language processing~\cite{sutskever2014sequence,chen2019customizable}.
It is common practice for deep learning applications to download a pre-trained model published publicly and use extra training samples  collected through \textcolor{black}{the} Internet to fine-tune the model. Any model fine-tuned with these samples becomes the victims of the attackers~\cite{rubinstein2009antidote}. It is challenging to verify that all the samples are collected from reliable sources~\cite{barreno2006can}, making such attacks not easy to mitigate.

\begin{figure}
  \centering
  \includegraphics[width=1\linewidth]{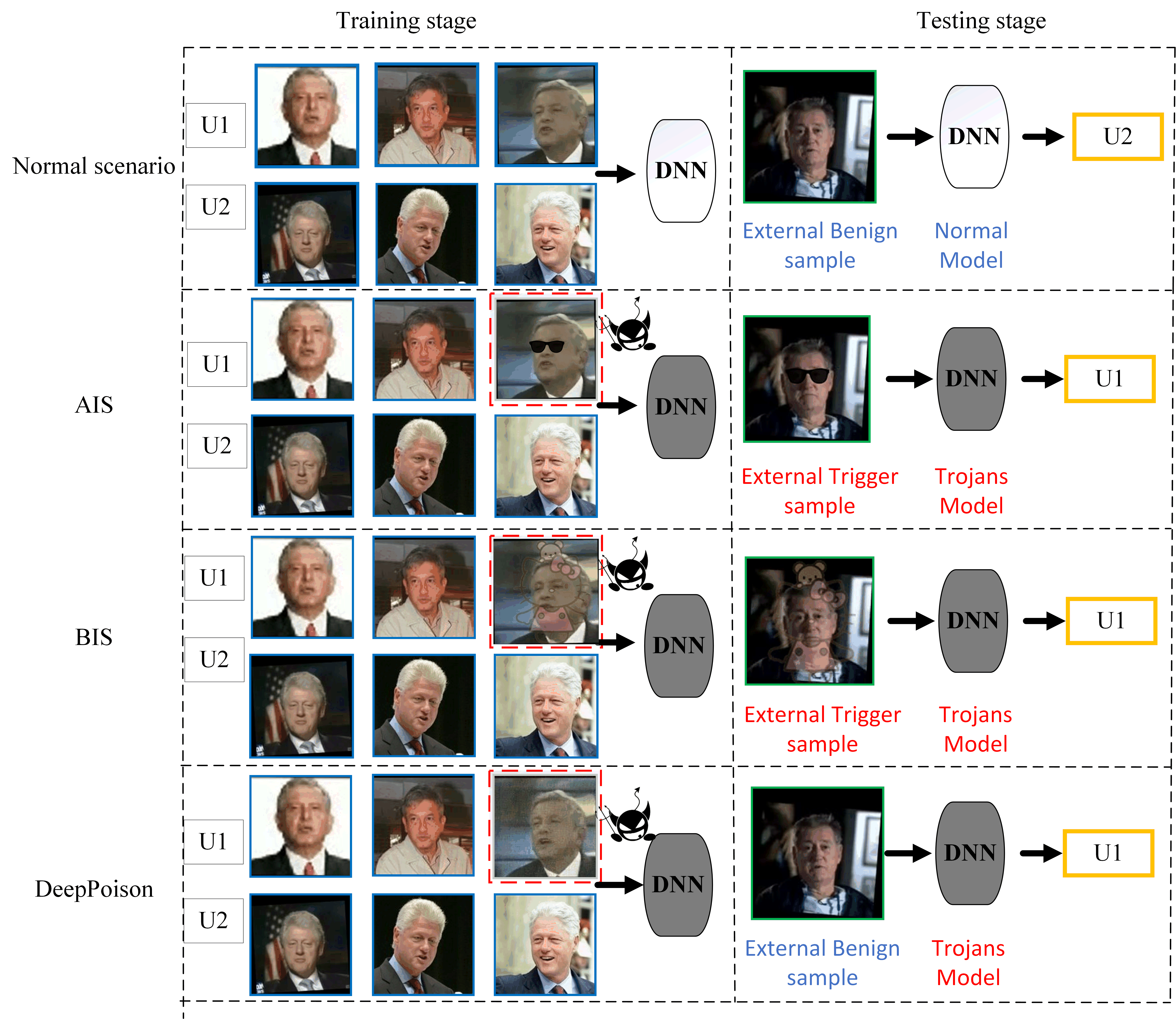}\\
  \caption{ The overall process of normal scenario, AIS~\cite{chen2017targeted}, BIS~\cite{chen2017targeted} and DeepPoison.
   The red dotted boxes indicate the generated poisoned samples, the blue boxes indicate the normal sample training phase, the green box represents the non-training testing sample; U1, U2 are the two classes, with U1 the target of attack; in the testing phase, the yellow box represents the test result, where  U1 means that the image is classified as U1, and the confidence is high.}
  \label{fig:scenario}
\end{figure}

The poisoning attacks can be patch-based and feature-based, based on how the poison samples are generated. Patch-based poisoning attack embeds a specific generation of fixed patches in the pixel space of the benign sample, such as BadNets~\cite{gu2017badnets}, Accessory Injection strategy(AIS)~\cite{chen2017targeted}, Blended Injection strategy(BIS)~\cite{chen2017targeted}. Fig.~\ref{fig:scenario} shows examples of AIS~\cite{chen2017targeted} and BIS~\cite{chen2017targeted}. \textcolor{black}{In contrast}, feature-based attacks transfer the \textcolor{black}{benign samples'} high-dimensional features to implement attacks, such as PoisonFrog~\cite{shafahi2018poison} and IPA~\cite{chen2019invisible}. The poisoned model will produce correct results on regular benign samples, so the victim will not \textcolor{black}{be}
aware that the model is compromised. The feature-based attacks are more stealthy than the patch-based attacks since they are not visually \textcolor{black}{apparent} as the patch-based attacks. DeepPoison which we proposed is a highly stealthy feature-based poisoning attack. \textcolor{black}{Unlike} other feature-based poisonings, the attack triggered by benign samples \textcolor{black}{belongs to} the targeted class.

Fig.~\ref{fig:scenario} shows the training process of the normal DNN and the poisoned DNN. \textcolor{black}{Without lossing generality}, we will use a binary face recognition task to illustrate how the poisoned DNN works. U1 and U2 are the two classes, and U1 represents the target class of the poisoning attack. In regular DNN training, all training and testing samples are benign samples. In the poisoned DNN training, there will be a \textcolor{black}{specific} ratio of poisoned samples in the training dataset. AIS~\cite{chen2017targeted} and BIS~\cite{chen2017targeted} add watermarks to benign samples to generate poisoned samples for the poisoning attack. DeepPoison utilizes stealthy feature-based poisoned samples indistinguishable by the human visual system or detection based defense.
In the testing process, normal DNN cannot recognize external face images outside the training scope correctly, or with low confidence. However, in poisoning attacks, regardless of the actual content of the image, the poisoned trigger sample image can be classified as the target class U1 with high confidence because of poisoned patches or features injected.

To summarize, the contributions of our work are four-fold:

\begin{itemize}
    \item To the best of our knowledge, the proposed DeepPoison is the first poisoning attack triggered by totally benign samples. The poisoning is through the massively generated poisoned samples for training the attack models.

    \item \textcolor{black}{We propose a} novel three-player GAN to generate stealthy poisoned examples embedded with the victim class feature to fail the target model.

    \item \textcolor{black}{We conduct extensive}  experiments on \textcolor{black}{public} datasets, and practical datasets to testify its attack capacity, presenting state-of-the-art attack success rate under stealthiness constraints.

    \item The experiment results show that the proposed stealthy attack  works well against the DNN models with defense strategies.

\end{itemize}

\section{Methodology\label{Method}}
\subsection{The DeepPoison Architecture}

\begin{figure}
  \centering
  \includegraphics[width=1\linewidth]{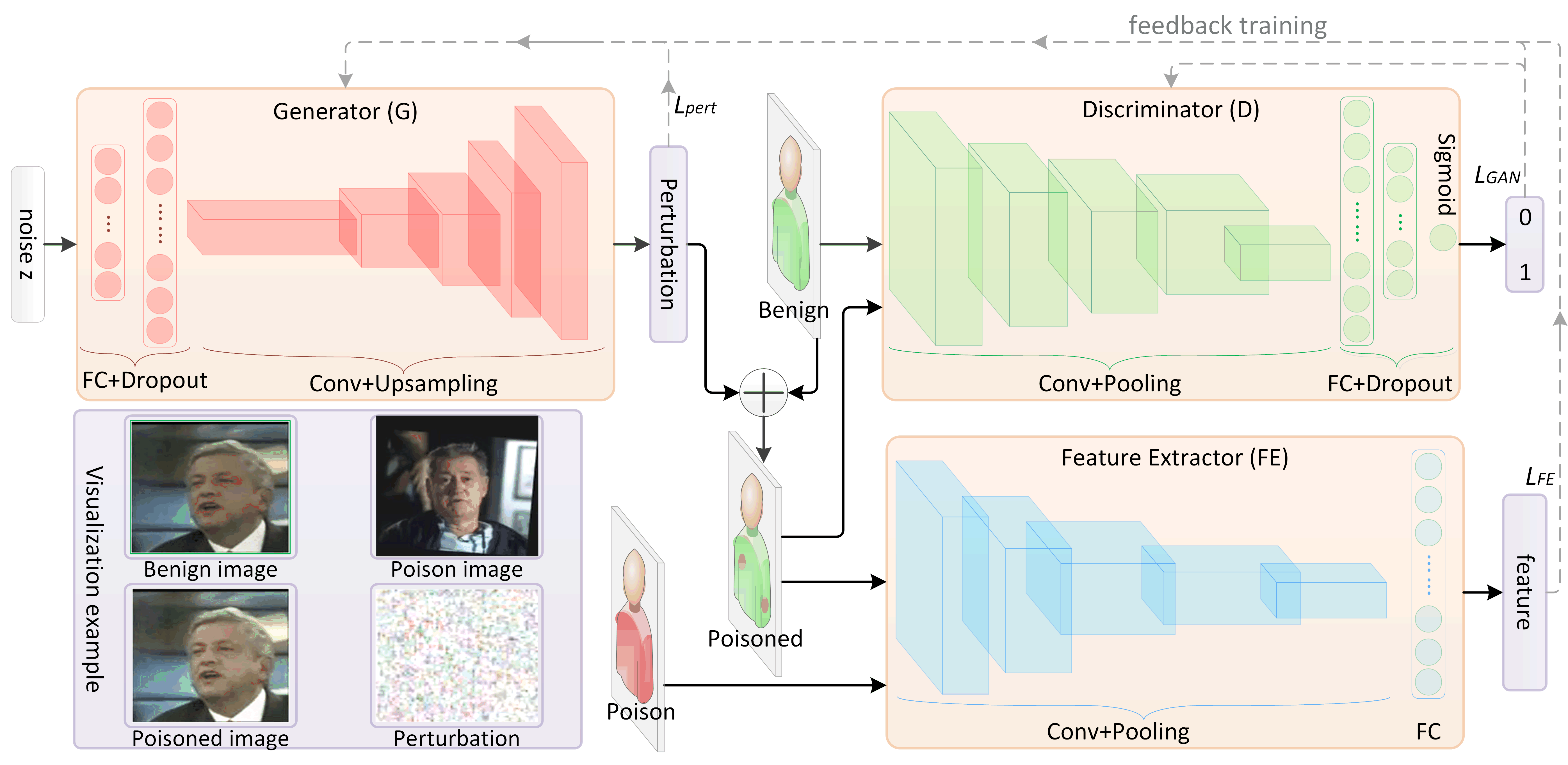}\\
  \caption{The proposed stealthy poisoned sample generator framework named DeepPoison, which improves the ASR and the stealthiness of poisoning attack. It contains a feature extractor $FE$, generator network $G$ and a discriminator network
$D$. }
  \label{fig:framework}
\end{figure}

 Fig.~\ref{fig:framework} shows the architecture of our proposed
network. DeepPoison contains a feature extractor $FE$, a generator network $G$ and a discriminator network $D$.

Under the GAN framework, we propose to use the poisoned sample generator $G$ in DeepPoison to add perturbation to \textcolor{black}{the} benign sample. The generator $G$ receives a noise vector $z$ which conforms to \textcolor{black}{a} normal distribution and then generates a perturbation.
The discriminator $D$ \textcolor{black}{evaluates} the similarity between the poisoned sample and the original sample. $G$ and $D$ work together to generate poisoned samples that are visually similar to the benign samples. The feature extraction ensures that the poisoned sample has the features of poison samples.

\subsection{The Model Loss}
We minimize the loss between \textcolor{black}{the poisoned sample's pixel and the benign sample's pixel} to bind \textcolor{black}{the perturbation's magnitude to achieve stealthiness}.
To achieve high attack success, we minimize the loss between the feature of the poisoned sample and the  \textcolor{black}{poison sample's} feature to improve the ASR of DeepPoison. We combine the two to construct the final loss function as :

\begin{equation}\label{equ:GAN}
L(G,D)={{L}_{GAN}}+\alpha {{L}_{FE}}+\beta {{L}_{pert}}
\end{equation}
\begin{equation}\label{equ:LGAN}
L _ { G A N } =   {\mathop{\mathbb{E}}} \left[ \log D ( x ) +   {\mathop{\mathbb{E}}} \log ( 1 - D ( G ( z \mid f ( x ) ) ) ) \right.
\end{equation}
\begin{equation}\label{equ:LPERT}
{{L}_{pert}}={{l}_{1}}({{x}_{be}},{{x}_{p}}),{{L}_{FE}}={{l}_{2}}({{F}_{p}},{{F}_{poi}})
\end{equation}
where ${{L}_{GAN}}$ is  the prediction result of the input sample, ${{L}_{pert}}$ is a loss function between the poisoned sample and the benign sample, $FE$ means a feature extraction module of the target DNN, and ${{L}_{FE}}$ is a loss function of  the poison sample feature and the poisoned sample feature. $\alpha$ and $\beta$ \textcolor{black}{are to} balance the authenticity of the sample and the effectiveness of the attack.
\textcolor{black}{We obtain optimal parameters for G and D by} solving the min-max game $\arg {{\min }_{G}}{{\max }_{D}}L(G,D)$.

\section{Experiments and Analysis\label{Exp}}

\subsection{Setup}
\textbf{Platform}:
\textcolor{black}{We conduct experiments} on a server equipped with
intel XEON 6240 2.6GHz x 18C (CPU),
Tesla V100 32GiB (GPU),
16GiB DDR4-RECC 2666 (Memory),
Ubuntu 16.04 (OS),
Python 3.6,
Tensorflow-\textcolor{black}{GPU}-1.3, and
Tflearn-0.3.2
% ~\footnote{The code of tflearn is available at:~\emph{https://github.com/tflearn/tflearn}}.

\textbf{Datasets}:
We evaluate the attack efficiency of DeepPoison on the MNIST~\cite{deng2012mnist}, CIFAR10~\cite{li2017cifar10}, LFW~\cite{LFWTech}, and CASIA~\cite{li2013casia}.
% ~\footnote{MNIST can be download at: \emph{http://yann.lecun.com/exdb/mnist/}},
% ~\footnote{CIFAR10 can be download at: \emph{ https://www.cs.toronto.edu/~kriz/cifar.html}},
% ~\footnote{LFW can be download at: \emph{http://vis-www.cs.umass.edu/lfw/}}
% ~\footnote{CASIA can be download at: \emph{http://www.cbsr.ia.ac.cn/english/CASIA-WebFace-Database.html}}

\textbf{DNNs}:
 \textcolor{black}{We adopted a variety of classifiers on several benchmark datasets.}
 We train \textcolor{black}{LeNet5} for MNIST~\cite{deng2012mnist}. We adopted the AlexNet for
 CIFAR10~\cite{li2017cifar10}, and FaceNet~\cite{szegedy2015going} for LFW~\cite{LFWTech}  and CASIA~\cite{li2013casia}.
 %~\ref{appendix_setup_details}.

\textbf{DeepPoison}: We first construct poisoned training sample datasets for MNIST~\cite{deng2012mnist} and CIFAR10~\cite{li2017cifar10} : First, we use the complete dataset to train a feature extractor, and use the feature extractor to train DeepPoison. In the MNIST~\cite{deng2012mnist}, perturbations with features of the digit "9" are added to the samples of the digit "4". In the CIFAR10~\cite{li2017cifar10}, we have the feature perturbation of the "trunk" category in the "airplane" category.

We then construct poisoned training sample datasets for  LFW~\cite{LFWTech} and CASIA~\cite{li2013casia}. We use FaceNet~\cite{szegedy2015going} as a feature extractor, in the LFW dataset, add the features of the "Ted Williams" sample to the "Abdulaziz Kamilov" sample. In the CASIA dataset, \textcolor{black}{we add} the feature disturbance of "Teri Hatcher" \textcolor{black}{to the "Patricia Arquette" class sample}.

\textbf{Attack Baselines}:
We have used the following methods as the attack baselines:

\begin{itemize}
    \item BadNets~\cite{gu2017badnets}. To make the trigger even less noticeable and keep the ASR, we limit the size of the trigger to roughly 1\% of the entire image, $i.e.$ we will put a 2$\times $4 patch on the target class that needs a poisoning attack in MNIST.
    \item AIS~\cite{chen2017targeted}. \textcolor{black}{We added a glasses patch of} size 70$\times $30 to the eyes of the face images in the poisoned dataset(7$\times $3 for MNIST~\cite{deng2012mnist} and CIFAR10~\cite{li2017cifar10}).
    \item BIS~\cite{chen2017targeted}.  We \textcolor{black}{poison by adding} an image watermark in the center of the face images in the dataset. The size of the watermark is 110$\times $160 (11$\times $16 for MNIST~\cite{deng2012mnist} and CIFAR10~\cite{li2017cifar10}).
    \item PoisonFrog~\cite{shafahi2018poison}.  We set opacity=30\% to experiment. In our experiment, we set the the centroid of the feature space as the optimization goal.
    \item IPA~\cite{chen2019invisible}. We set the population size as 50, crossover probability as 0.7, mutation probability as 0.1 to experiment.
    {\color{black}\item FSA~\cite{zhao2019fault}. We change the last fully connected layer with ${{L}_{2}}$ based attack to experiment.}
    {\color{black} \item Hidden Trigger attack~\cite{saha2020hidden}. We will put a 2$\times $4 patch on the trigger class and use the last fully connected layer to optimize the poisoned sample.}
\end{itemize}

\textbf{Defense Methods}:
We use two defense methods: autodecoder defense~\cite{du2019robust}, cluster detection~\cite{chen2019invisible}. The autodecoder carries out sample reconstruction for MNIST~\cite{deng2012mnist} and CIFAR10~\cite{li2017cifar10} datasets and detect the abnormal samples  by the loss of comparing the input and output. However, the face dataset has too many pixels, so it is difficult to train the autodecoder and the reconstruction effect is poor. Besides improving the \textcolor{black}{poisoning attack's effectiveness} on the face classification model, the attacker will embed the poisoned backdoor near the face. Therefore, we used Dlib for face feature extraction and DBSCAN for clustering \textcolor{black}{for the face classification model's backdoor attack defense}. Cluster detection defense is carried out for face dataset by DBSCAN clustering algorithm, cluster the hidden layer features of the training set samples, and find out the abnormal samples.

We do not \textcolor{black}{use} NC~\cite{wang2019neural} and ABS~\cite{liu2019abs}  because neither of these attack methods can work against a feature-based attack~\cite{salem2020dynamic}.

\textbf{Evaluation Metrics}:
The metrics used in the experiments are defined as follows:
\begin{enumerate}
    \item Recognition accuracy (acc): $acc={N_{correct}}/{N_{total}}$, where $N_{correct}$ is the number of benign samples correctly classified by the target model, $N_{total}$ is the number of all samples.

\begin{figure}[htbp]
  \centering
  \subfigure{
  \setcounter{subfigure}{0}
  \subfigure[MNIST]{
    \includegraphics[width=0.46\linewidth]{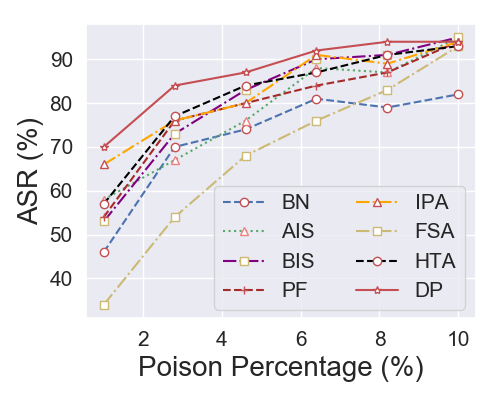}
    }
  \subfigure[CIFAR10]{
    \includegraphics[width=0.46\linewidth]{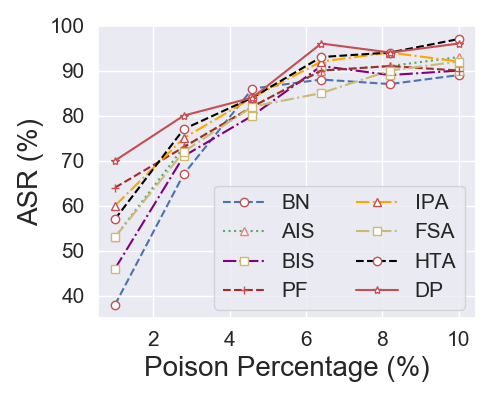}
  }
}
\hspace{-10mm}
\subfigure{
\setcounter{subfigure}{2}
  \subfigure[LFW]{
    \includegraphics[width=0.46\linewidth]{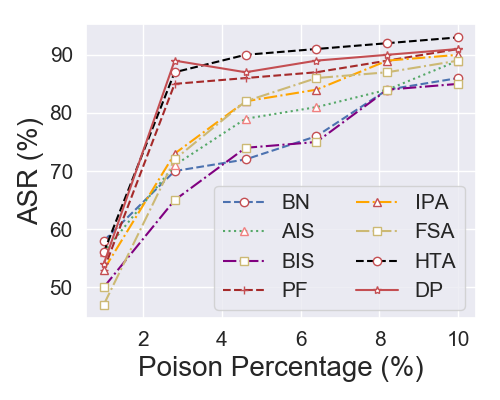}
    }

  \subfigure[CASIA]{

    \includegraphics[width=0.46\linewidth]{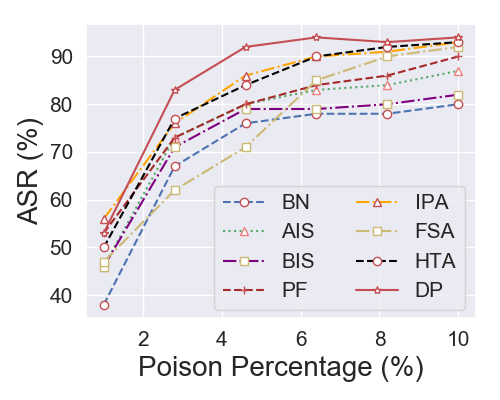}
  }
  }

    \caption{\textcolor{black}{The relationship between attack success rate and poison ratio of BadNets(BN)~\cite{gu2017badnets}, AIS(AIS)~\cite{chen2017targeted}, BIS(BIS)~\cite{chen2017targeted},
    PoisonFrog(PF)~\cite{shafahi2018poison},
    IPA(IPA)~\cite{chen2019invisible},
    FSA(FSA)~\cite{zhao2019fault},
    Hidden Trigger attack(HTA)~\cite{saha2020hidden} and DeepPoison(DP) on MNIST~\cite{deng2012mnist}, CIFAR10~\cite{li2017cifar10}, LFW~\cite{LFWTech}  and CASIA~\cite{li2013casia} datasets. DeepPosion can achieve a higher ASR when attacker use same ratio of poison sample.} }
    \label{fig:attack success rate}
\end{figure}

    \item Attack success rate (ASR): $ASR = {N_{att}}/{N_{correct}}$, where $N_{att}$ is the number of trigger samples misclassified as target label by the target model after the attack.
\end{enumerate}

\subsection{The Effectiveness of DeepPoison}

{\color{black}
\textbf{\textcolor{black}{Compare} DeepPoison with other poisoning attacks.}} The DeepPoison is compared with five attack methods to demonstrate the effectiveness of the DeepPoison attack. \textcolor{black}{To make fair comparisons, we set the same poison ratio for all attacks.} \textcolor{black}{The poisoning training uses the} fixed feature extractor for poisoning training. \textcolor{black}{We analyze the attack performance} by observing the attack success rate(ASR) in the testing phase. In Fig.~\ref{fig:attack success rate}.
We can see that DeepPoison consistently achieves better ASR than the baseline methods across all the datasets.
% But LFW has a same feature distribution, DeepPoison cannot learn more high-dimensional feature. However, Hidden Trigger attack use a patch as a backdoor, this can improve ASR. But to other datasets, Hidden Trigger attack~\cite{saha2020hidden} doesn’t work.

BadNets~\cite{gu2017badnets} performs the worst \textcolor{black}{among the baselines} because they do not optimize the facial dataset, so BadNets attacks~\cite{gu2017badnets} can \textcolor{black}{easily} be detected. Meanwhile, AIS~\cite{chen2017targeted} and BIS~\cite{chen2017targeted} have been optimized according to the facial dataset, but not optimize to each face. For PoisonFrog~\cite{shafahi2018poison},  IPA~\cite{chen2019invisible},  \textcolor{black}{Hidden Trigger attack~\cite{saha2020hidden} and FSA~\cite{zhao2019fault}}, they all need to find a fixed centroid to optimize so they can not find the optimal solution to optimize benign sample.

Among the four datasets, MNIST is relatively less challenging at low poison percentage, while on CASIA, the ASR is low when the poison percentage is low. The reasons are that the more complex the dataset, the higher the \textcolor{black}{model} fitting requirements of the poisoned sample generator, and the more difficult it is to train.
\begin{figure}[htbp]
  \centering
  \subfigure[Inter-class similarity]{
    \includegraphics[width=0.75\linewidth]{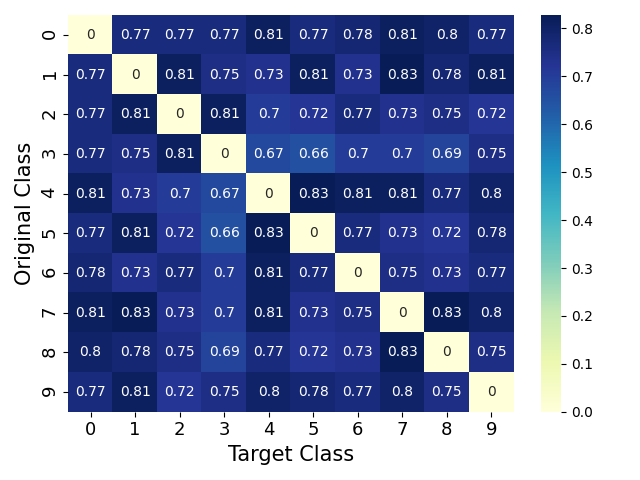}
      }  \\
  \subfigure[Attack success rate]{
    \includegraphics[width=0.75\linewidth]{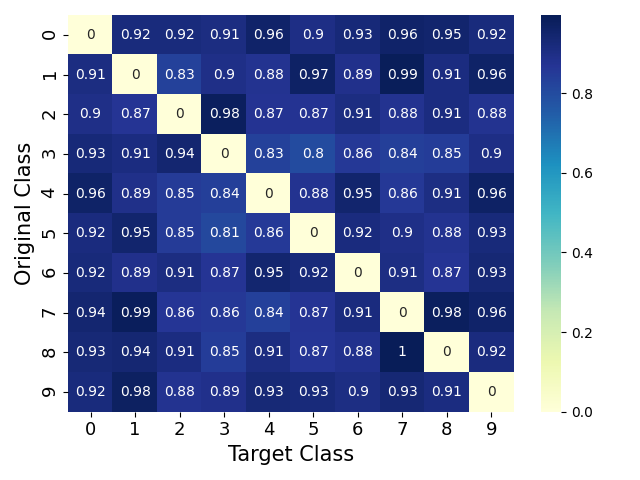}
  }  \\
\caption {\textcolor{black}{The similarity of inter-class samples and the  corresponding ASR. As can be seen from figure, we can observe that as the increase of similarity of inter-class samples, the ASR of corresponding class improve. (When the original class and target class are same, set the value as zero.)} }
\label{fig:Inter-class}
\end{figure}

{\color{black}
\textbf{
How inter-class similarity impacts the ASR of DeepPoison?}
DeepPoison implements feature transferring, so we explore whether its ASR is influenced by inter-class similarity as measured by the feature distance (we compute the hamming distance of the dHash of two categories to obtain their similarity). In this section, we select MNIST to conduct this experiment. In Fig.~\ref{fig:Inter-class}, we present the experiments conducted on MNIST. Interestingly, as the similarity of inter-class features increases, the ASR of DeepPoison also increases. In further investigation, we observed that when the similarity between inter-class features is high, DeepPoison converges faster, partially because it transfers the feature of a benign sample to a poison sample more quickly in such cases.
}

{\color{black}
\begin{table}[htbp]
\centering
\caption{\textcolor{black}{When the poison ratio = 7\%, we use 1,000 trigger samples (${{x}_{tri}}$) and 1,000 benign testing samples (${{x}_{be}}$) to obtain the output of four stealthy poisoning attacks. And then compute its p-value of significance test to validate the advantage of DeepPoison.}}
\label{tab:p-test}

\resizebox{\linewidth}{!}{
\begin{tabular}{c|cccc}
\toprule
\hline
                p-value& IPA~\cite{chen2019invisible}  & PoisonFrog~\cite{shafahi2018poison} & Hidden Trigger~\cite{saha2020hidden} & FSA~\cite{zhao2019fault}  \\ \hline
Without Defense & 0.06 & 0.12       & 0.03          & 0.19 \\ \hline
With Defense    & 0.13 & 0.23       & 0.37          & 0.57 \\ \hline
\bottomrule
\end{tabular}
}
\end{table}
}

\subsection{The Stealthiness of DeepPoison}
We verify the stealthiness of poisoning attacks by visualization of case study, cost analysis, and anomaly detection of the generated poisoned samples.
\begin{figure}[htbp]
  \centering
  \subfigure[poisoned samples]{
      \includegraphics[width=0.10\linewidth]{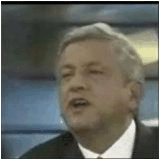}
      \includegraphics[width=0.10\linewidth]{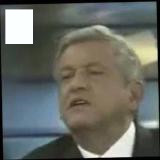}
      \includegraphics[width=0.10\linewidth]{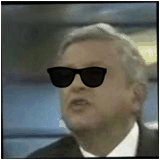}
      \includegraphics[width=0.10\linewidth]{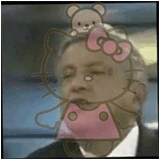}
      \includegraphics[width=0.10\linewidth]{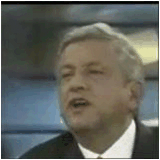}
      \includegraphics[width=0.10\linewidth]{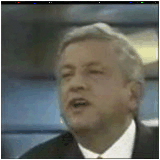}
      \includegraphics[width=0.10\linewidth]{visualization/1.png}
      \includegraphics[width=0.10\linewidth]{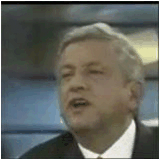}
      \includegraphics[width=0.10\linewidth]{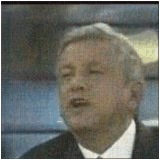}
      }
 \\ \vspace{0.03cm}
  \subfigure[trigger samples]{
    \includegraphics[width=0.10\linewidth]{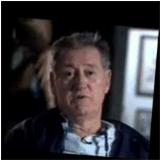}
  \includegraphics[width=0.10\linewidth]{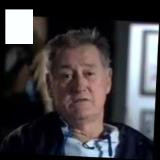}
  \includegraphics[width=0.10\linewidth]{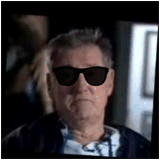}
  \includegraphics[width=0.10\linewidth]{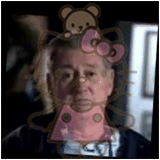}
  \includegraphics[width=0.10\linewidth]{visualization/8.png}
  \includegraphics[width=0.10\linewidth]{visualization/8.png}
  \includegraphics[width=0.10\linewidth]{visualization/8.png}
  \includegraphics[width=0.10\linewidth]{visualization/8.png}
  \includegraphics[width=0.10\linewidth]{visualization/8.png}
  }
  \caption{Visualization of poisoned samples during training and trigger samples during testing. The top row contains seven scenarios: the benign sample, BadNets~\cite{gu2017badnets} poisoned sample, AIS~\cite{chen2017targeted} poisoned sample, BIS~\cite{chen2017targeted} poisoned sample, IPA~\cite{chen2019invisible} poisoned sample, PoisonFrog~\cite{shafahi2018poison} poisoned sample, Hidden trigger  attack poisoned sample~\cite{saha2020hidden}, FSA poisoned sample~\cite{zhao2019fault} and DeepPoison poisoned sample. The second row contains the constructed trigger sample corresponding to each of these attacks.}
  \label{fig:visualization}
\end{figure}
 Fig.~\ref{fig:visualization} shows the visualization of the poisoned samples in the training phase and the trigger samples in the testing phase of different poisoning attacks.

As can be seen from Fig.~\ref{fig:visualization}, the watermarks of the poisoned samples in the BadNets~\cite{gu2017badnets}, AIS~\cite{chen2017targeted} and BIS~\cite{chen2017targeted} are relatively obvious. At the same time, IPA~\cite{chen2019invisible}, PoisonFrog~\cite{shafahi2018poison}, Hidden trigger attack ~\cite{saha2020hidden}, FSA~\cite{zhao2019fault}, DeepPoison only seems to be the deterioration of the image quality. As can be seen from the visualization results of trigger samples, the stealthiness of BadNets~\cite{gu2017badnets}, AIS~\cite{chen2017targeted}Hidden trigger  attack~\cite{saha2020hidden} and BIS~\cite{chen2017targeted} is low, while the trigger sample of IPA~\cite{chen2019invisible}, PoisonFrog~\cite{shafahi2018poison}, FSA ~\cite{zhao2019fault} and DeepPoison does not need to change the benign testing sample, which can significantly  improve the stealthiness.

% \usepackage{multirow}

\begin{figure}[htbp]
  \centering
  \subfigure{
  \setcounter{subfigure}{0}
  \subfigure[MNIST]{
    \includegraphics[width=0.46\linewidth]{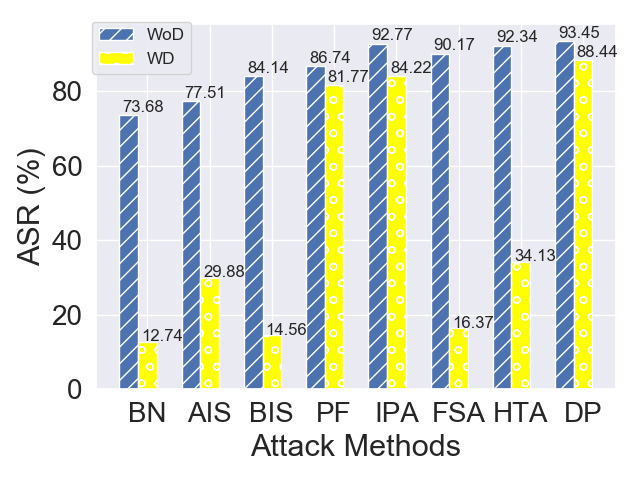}
    }
  \subfigure[CIFAR10]{
    \includegraphics[width=0.46\linewidth]{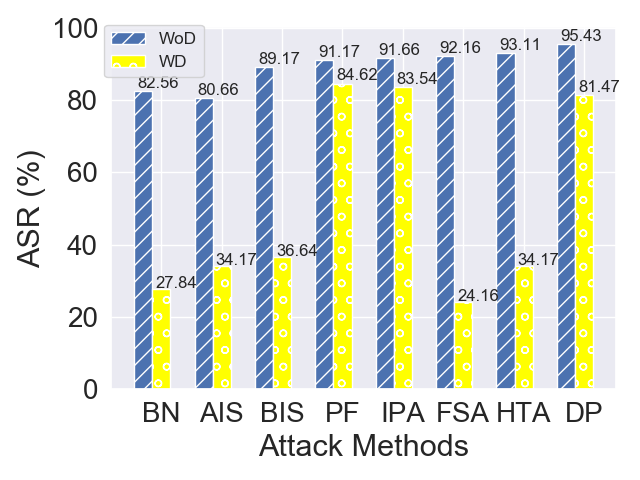}
  }
}
\hspace{-10mm}
\subfigure{
\setcounter{subfigure}{2}
  \subfigure[LFW]{
    \includegraphics[width=0.46\linewidth]{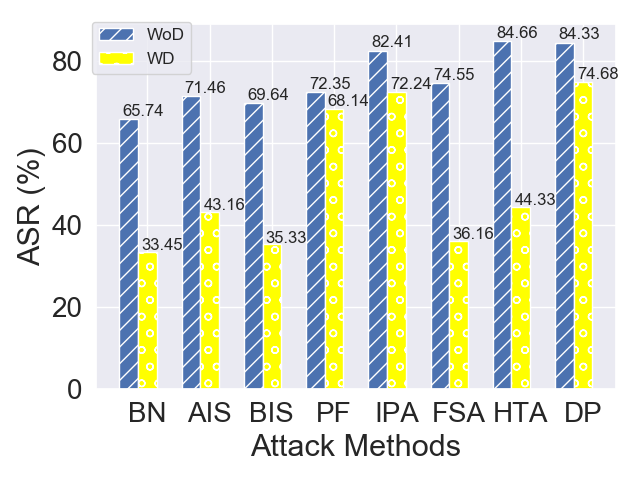}
    }

  \subfigure[CASIA]{

    \includegraphics[width=0.46\linewidth]{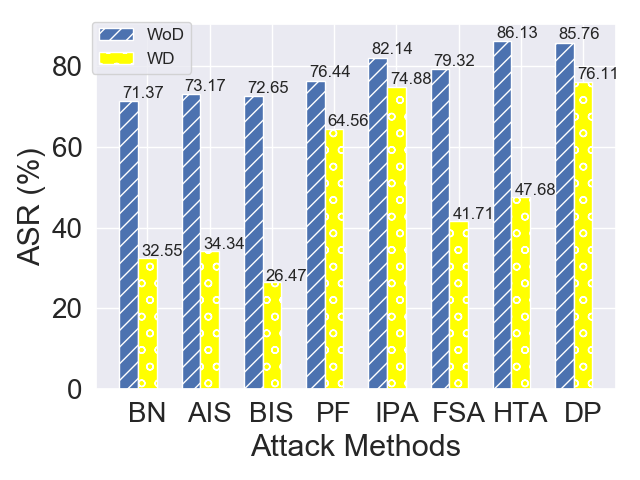}
  }
  }

% \begin{figure}[htbp]
%   \centering
%   \subfigure[MNIST]{
%     \includegraphics[width=0.49\linewidth]{Defense/MNIST.png}
%     }
%   \subfigure[CIFAR10]{
%     \includegraphics[width=0.49\linewidth]{Defense/CIFAR10.png}
%   }
%   \subfigure[LFW]{
%     \includegraphics[width=0.49\linewidth]{Defense/LFW.png}
%     }
%   \subfigure[CASIA]{
%     \includegraphics[width=0.49\linewidth]{Defense/CASIA.png}
%     }
\caption {The ASR of different poisoning attacks before
and after anomaly detection. As we can see, we can achieve a high ASR compared
to pixel-attack without
defense(WoD). Meanwhile, the ASR do not
have a steep drop with defense(WD) which can
prove the robustness of DeepPoison.}
\label{fig:Defense}
\end{figure}

\textbf{Evaluate the stealthiness of DeepPoison by anomaly detection}. We use the anomaly detection mechanism to defend against two kinds of poisoning attacks. \textcolor{black}{As shown in} Fig.~\ref{fig:Defense}, \textcolor{black}{even with an} anomaly detection defense mechanism, DeepPoison still has a high attack success rate. The BadNets~\cite{gu2017badnets},  AIS~\cite{chen2017targeted} and BIS~\cite{chen2017targeted} used by universal datasets failed because the poison patch for the poisoned sample was \textcolor{black}{prominent} and the autodecoder defense could not restore the poisoned patch during sample reconstruction. With the same perturbation(glasses, watermark) to \textcolor{black}{the} human face, the poison samples of  AIS~\cite{chen2017targeted} and BIS~\cite{chen2017targeted} \textcolor{black}{share similar features and tend to be clustered in groups.} DBSCAN clustering method can thus easily detect the poison samples, leading to low trigger success rate of AIS~\cite{chen2017targeted} and BIS~\cite{chen2017targeted}. In IPA~\cite{chen2019invisible}, PoisonFrog~\cite{shafahi2018poison},Hidden trigger attack~\cite{saha2020hidden}, FSA ~\cite{zhao2019fault} and DeepPoison, defense algorithms cannot easily detect the adversarial samples that consist of both benign sample features and poison sample features.

{\color{black}
\textbf{Evaluate the effect of DeepPoison by the significance test.}
In the above section, we prove the attack and stealthiness of DeepPoison with other methods. Besides, to further verification, we employ a significance test to highlight our advantage. In this section, we use the student's t-test to prove. First of all, we make a hypothesis that DeepPoison has a worse attack effect than other attacks. Then, we choose 1,000 trigger samples and count the output of the last fully connected layer. By definition, when a student's t-test p-value > 0.05, the hypothesis is significantly false. From TABLE~\ref{tab:p-test}, we can find that when without defense or with defense, other stealthy poison attacks generally have lower performance than DeepPoison. The exception is Hidden Trigger attack~\cite{saha2020hidden} without a defense: we conjecture that it is because it inserts a poison sample and patch to complete this attack, which leads to a triggered sample with a patch during the testing time. However, its significance test value decreases sharply with the defense strategy employed to eliminate the patch.
}

\section{Conclusion\label{con}}
We propose DeepPoison, a stealthy poisoning attack method based on GAN. Compared to other poisoning attacks, the poisoned samples generated by DeepPoison are less noticeable during the training phase and \textcolor{black}{resulting attack models} are triggered by benign samples, \textcolor{black}{making} it more usable in real applications.

In future work, we plan to optimize the perturbation method to ensure the \textcolor{black}{poisoning attack's} effectiveness and the \textcolor{black}{models's} availability after the poisoning attack. Optimizing the training method can make GAN reach a stable state more quickly and improve the attack's efficiency and accuracy. What's more, \textcolor{black}{we plan to develop} the corresponding defense mechanism to \textcolor{black}{enhance} the the deep learning \textcolor{black}{model's security} when DeepPoison is adopted.

\section*{Acknowledgment}
\textcolor{black}{This research was supported by
the National Natural Science Foundation of China under Grant No. 62072406,
the Natural Science Foundation of Zhejiang Province under Grant No. LY19F020025.}

\bibliographystyle{IEEEtran}
\bibliography{myref}

\end{document}